\documentclass[aps,prb,twocolumn,superscriptaddress]{revtex4-1}

\usepackage{graphicx}
\usepackage[version=4]{mhchem}
\usepackage{siunitx}
\usepackage[table]{xcolor}
\usepackage{tabularx}
\usepackage{multirow}
\usepackage{amsfonts}
\usepackage{amssymb}
\usepackage[]{amsmath}
\usepackage[bookmarks=true,
			bookmarksnumbered=true,	
			colorlinks=true,
			urlcolor=black,
			linkcolor=black,
			citecolor=blue]{hyperref}
\usepackage{changes}

\newcommand*{\sect}{Section\,}
\newcommand*{\sects}{Sections\,}

\newcommand*{\fig}{Fig.\,}
\newcommand*{\figs}{Figs.\,}
\newcommand*{\eq}[1]{Eq.\,(#1)}
\newcommand*{\tab}{Tab.\,}
\newcommand*{\reftaken}{Ref.\,}

\newcommand*{\etal}{\emph{et~al. }}

\newcommand*{\etalk}{\emph{et~al.}}
\newcommand*{\ie}{{i.e., }}

\newcommand*{\captionheader}[1]{{#1}}
\newcommand*{\eg}{e.g., }
\newcommand*{\cf}{cf.\,}

\newcommand*{\citen}{}
\DeclareRobustCommand*{\citen}[1]{%
  \begingroup
    \romannumeral-`\x 
    \setcitestyle{square,numbers}%
    \cite{#1}%
  \endgroup
}

\begin{document}
\title{Evolution of the spin dynamics during freezing in the spin-glass \ce{Fe_{x}Cr_{1-x}}}

\author{S. S{\"a}ubert}
\email[]{steffen.saeubert@gmail.com}

\affiliation{Physik-Department, Technische Universit{\"a}t M{\"u}nchen, D-85748 Garching, Germany}
\affiliation{Heinz Maier-Leibnitz Zentrum, Technische Universit{\"a}t M{\"u}nchen, D-85748 Garching, Germany}

\author{C. Franz}
\affiliation{Physik-Department, Technische Universit{\"a}t M{\"u}nchen, D-85748 Garching, Germany}
\affiliation{Heinz Maier-Leibnitz Zentrum, Technische Universit{\"a}t M{\"u}nchen, D-85748 Garching, Germany}
\affiliation{Jülich Centre for Neutron Science (JCNS) at Heinz Maier-Leibnitz Zentrum (MLZ), Forschungszentrum Jülich GmbH, Garching, Germany}

\author{J.K. Jochum}
\affiliation{Physik-Department, Technische Universit{\"a}t M{\"u}nchen, D-85748 Garching, Germany}
\affiliation{Heinz Maier-Leibnitz Zentrum, Technische Universit{\"a}t M{\"u}nchen, D-85748 Garching, Germany}

\author{G. Benka}
\affiliation{Physik-Department, Technische Universit{\"a}t M{\"u}nchen, D-85748 Garching, Germany}

\author{A. Bauer}
\affiliation{Physik-Department, Technische Universit{\"a}t M{\"u}nchen, D-85748 Garching, Germany}
\affiliation{Zentrum f\"ur QuantumEngineering (ZQE), Technische Universit\"at M\"unchen, D-85748 Garching, Germany}

\author{S.M. Shapiro}
\affiliation{Brookhaven National Laboratory, Department of Physics, Upton, NY 11973, USA}

\author{P. B{\"o}ni}
\affiliation{Physik-Department, Technische Universit{\"a}t M{\"u}nchen, D-85748 Garching, Germany}

\author{C. Pfleiderer}
\affiliation{Physik-Department, Technische Universit{\"a}t M{\"u}nchen, D-85748 Garching, Germany}
 \affiliation{Zentrum f\"ur QuantumEngineering (ZQE), Technische Universit\"at M\"unchen, D-85748 Garching, Germany}
 \affiliation{Munich Center for Quantum Science and Technology (MCQST), Technische Universit\"at M\"unchen, D-85748 Garching, Germany}

\date{\today}

\begin{abstract}
In the iron--chromium system, Fe$_{x}$Cr$_{1-x}$, a wide dome of spin-glass behavior emerges when the ferromagnetism of iron is suppressed and the antiferromagnetism of chromium emerges as a function of increasing iron content $x$. As both, the high-temperature state and the characteristic cluster size vary as a function of $x$, different regimes of spin-glass behavior may be compared in a single, isostructural material system. Here, we report a study of the spin dynamics across the freezing process into the spin-glass state for different iron concentrations ($x = 0.145$, $0.175$, $0.21$) using Modulation of IntEnsity with Zero Effort (MIEZE) spectroscopy. In the parameter range studied, the relaxation process observed experimentally may be described well in terms of a stretched exponential. In the reentrant cluster-glass regime, $x = 0.145$, this behavior persists up to high temperatures. In comparison, in the superparamagnetic regime, $x = 0.175$ and $x = 0.21$, a single relaxation time at elevated temperatures is observed. For all samples studied, the spin relaxation exhibits a momentum dependence consistent with a power law, providing evidence of a dispersive character of the spin relaxation.
\end{abstract}

\maketitle

\subsection{\label{sec:introduction}Introduction}

In spin-glasses, the combination of random site occupation and disorder with competing interactions, anisotropy, and frustration leads to a collective freezing of the spins in random orientations at the glass temperature $T_{\mathrm{g}}$\cite{Nordblad_2013, 2015_Mydosh_RepProgPhys}. 
The freezing process may involve a broad distribution of relaxation times, resulting from varying correlation lengths of the individual magnetic clusters. These clusters may vary strongly in size, ranging from individual spins in canonical spin-glasses, over interacting clusters of spins in cluster glasses, to collectively behaving domains in superparamagnetic systems\cite{McCloy2019}. 
In dilute spin-glasses, the perhaps most comprehensively studied class of glassy magnetic systems~\cite{1979_Mezei_JMagnMagnMater, 1981_Murani_PhysicaB+C, 1981_Murani_JMagnMagnMater, 1982_Mezei_JApplPhys, 1983_Mezei_JMagnMagnMater, 1984_Heffner_PhysRevB, 1985_Shapiro_JApplPhys, 2003_Pappas_PhysRevB, 2018_Wagner_QuantumBeamSci}, the relaxation times of the spin freezing exhibit no dispersion due to the short range of the interactions.

Seminal studies of the spin-glass behavior in \ce{Cu_{1-x}Mn_{x}} ($x\,=\,0.1$, $0.16$, and $0.35$) as well as \ce{Au_{1-x}Fe_{x}} ($x\,=\,0.14$) using neutron spin echo (NSE) spectroscopy~\cite{2003_Pappas_PhysRevB,2009_Pickup_PhysRevLett} suggested non-exponential relaxation. To go beyond an account of the spin relaxation in terms of a stretched exponential, which cannot distinguish between a distribution of parallel relaxation channels and hierarchical relaxation comprising intercluster and intracluster processes, the Weron model~\cite{1991_Weron_JPhysCondensMatter} was found to provide a universal description. This raises the question for key characteristics of the spin relaxation when the concentration of magnetic atoms is increased to form cluster-glass or superparamagnetic systems. Moreover, reports of a momentum independent quasielastic peak in the spin-glass regime~\cite{1980_Fincher_PhysRevLett, 1981_Shapiro_PhysRevB, 1983_Motoya_PhysRevB, aeppli_spin_1984, wicksted_investigation_1984, raymond_spin_1997} contrast dispersive behavior that has been attributed to either the coexistence of ferromagnetism and spin-glass behavior~\cite{murani_spin_1977, wagner_dynamics_2004, pappas_spin_2007} or momentum-dependent dynamics of the spin-glass~\cite{1999_Bao_PhysRevLett}.

To advance these questions, we report a study of the spin-glass dynamics in \ce{Fe_{x}Cr_{1-x}}. When combining the itinerant-electron ferromagnet iron and the spin-density wave antiferromagnet chromium in the isostructural alloy \ce{Fe_{x}Cr_{1-x}}, the ferromagnetic transition temperature is suppressed and long-range spin-density wave order emerges with increasing $x$ for $x_c\,=\,0.15\,-\,0.17$~\cite{1994_Fawcett_RevModPhys, 1979_Strom-Olsen_JPhysFMetPhys, 2001_Stewart_RevModPhys, 2007_Lohneysen_RevModPhys, 2008_Broun_NatPhys, 2019_Mirebeau_PhysRevB}. 
A dome of spin-glass behavior is located at low temperatures in the vicinity of $x_c$~\cite{1963_Nevitt_JApplPhys, 1966_Arajs_JApplPhys, 1972_Mitchell_PhysRevB, 1975_Loegel_JPhysFMetPhys, 1967_Ishikawa_JPhysSocJpn, 1978_Burke_JPhysFMetPhys, 1978_Burke_JApplCryst, 1981_Shapiro_PhysRevB}. 
The dome extends from $x\,\sim\,0.10$ to $x\,\sim\,0.25$, reaching well into concentration regimes which exhibit ferromagnetic and antiferromagnetic order at higher temperatures, respectively~\cite{1983_Burke_JPhysFMetPhysb, 1983_Burke_JPhysFMetPhys, 1983_Burke_JPhysFMetPhysa, 2022_Benka_PhysRevMaterials}. Thus, for different $x$, the spin-glass behavior in \ce{Fe_{x}Cr_{1-x}} may emerge with decreasing temperature from ferromagnetic or antiferromagnetic order due to spin freezing~\cite{1990_Mirebeau_PhysRevB, 1993_Mydosh_Book, Motoya1983, Hennion1983}. Moreover, the ferromagnetically ordered clusters initially grow in size as a function of increasing $x$, changing the character of the spin-glass state from a cluster-glass to a superparamagnet~\cite{2010_coey, 2019_glasses, 2022_Benka_PhysRevMaterials}. As both iron and chromium display spin wave dispersions that are prototypical for ferro- and antiferromagnetism, respectively, a question concerns the existence and character of dispersive behavior in the spin-glass regime~\cite{1969_Collins_PhysRev, Perring_1991, 2017_Kindervater_PhysRevB,2019_Saubert_PhysRevB,1988_Fawcett_RevModPhys, Fisher_1972, Als-Nielsen_1969,lequien_reinvestigation_1988}. 

Quasielastic neutron scattering has been established as an indispensable tool in the study of spin relaxation processes.
For a wide range of relaxation times expected in spin-glasses, neutron spin-echo spectroscopy appears to be ideally suited, since the associated measurements of the intermediate scattering function $S(q, \tau)$, as opposed to the dynamic structure factor $S(q, \hbar \omega)$, allows to separate dynamic processes on very different time scales.
However, in conventional neutron spin-echo spectroscopy depolarizing samples or samples under depolarizing sample environment may only be measured at a high penalty in neutron flux \cite{MEZEI20019, 1986_Farago_PhysicaB+C}. 
On this note, Modulation of IntEnsity with Zero Effort\cite{1998_Besenbock_JNeutronRes, 1992_Gahler_PhysicaBCondensMatter} (MIEZE) is a spin-echo technique which permits measurements under depolarizing conditions for small momentum transfers. It is therefore ideally suited for the study of spin relaxation dynamics in nearly ferromagnetic and ferromagnetic systems. 

In the study reported in this paper, we used the longitudinal MIEZE technique to determine the spin dynamics in \ce{Fe_{x}Cr_{1-x}} as a function of temperature for three different compositions.
For the reentrant cluster-glass ($x\,=\,0.145$), a broad distribution of relaxation times is observed over the entire temperature range that may be described well in terms of a stretched exponential. For compositions in the superparamagnetic regime ($x\,=\,0.175$ and $x\,=\,0.21$), a broad distribution of relaxation times at low temperatures that may be described well in terms of a stretched exponential is contrasted by a single relaxation time at elevated temperatures. Our data recorded in the sample with $x\,=\,0.175$ are consistent with an earlier study using transverse NRSE and conventional NSE in the same sample~\cite{2018_Wagner_QuantumBeamSci}, which, however exhibited more scatter and covered a smaller dynamical range. 
For all three compositions we find dispersive behavior of the spin relaxation that follows a power-law dependence of the momentum, $q$, consistent with $\Gamma\,\,\propto q^z$, where the dynamical exponent $z$ decreases from $z\,\sim\,1.5$ to $z\,\sim\,1.0$ with increasing $x$. This behavior is loosely reminiscent of the quasielastic linewidth in \ce{Fe_{0.7}Al_{0.3}}~\cite{1999_Bao_PhysRevLett}. However, for \ce{Fe_{x}Cr_{1-x}} the origin of the dispersive behavior may, in principle, comprise a combination of different contributions.

Our paper is organized as follows.
In \sect\ref{sec:experiment}, specific aspects of the neutron scattering experiments and the crystal growth techniques are presented. In \sects \ref{sub:elastic-scattering-fecr} and \ref{sub:quasielastic-measurements-fecr} the elastic and quasielastic neutron scattering experiments are described. In \sect\ref{sec:discussion}, the implications of the experiments are discussed, followed by a summary of the findings of this study in \sect\ref{sec:conclusion}.

\subsection{\label{sec:experiment}Experimental Methods}

\begin{figure}[t]
    \centering
    \includegraphics[width=1.0\columnwidth]{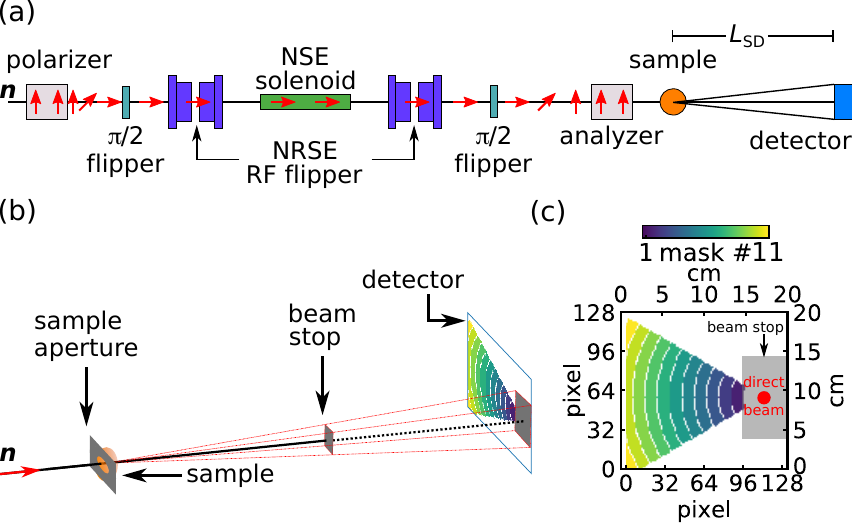}
    \caption{Longitudinal MIEZE setup used in our experiments\cite{2016_Krautloher_RevSciInstrum,2019_Saubert_PhysRevB}.  
    (a) Schematic depiction of the spectrometer. All spin manipulations are performed upstream to the sample, rendering the method insensitive to depolarizing conditions at the sample position. The red arrows indicate the direction of the magnetic guide fields. 
    (b) Schematic depiction of the neutron flight path through the spectrometer showing the sample aperture, sample, beam stop, and detector.
    (c) Detector segments used in the evaluation of the MIEZE scans. Segments represent parts of circular rings centered at the direct beam with an opening angle of \SI{60}{\degree} and a width of \SI{10}{pixels}. The grey shaded area represents the location of the beam stop.
    }
    \label{fig:fig_app_1_setup_masks}
\end{figure}

Our experiments were conducted at the beamline RESEDA \cite{2015_Franz_JLSRF,2019_Franz_NuclInstrumMethodsPhysRes} at the Heinz Maier-Leibnitz Zentrum using the longitudinal MIEZE option \cite{2016_Krautloher_RevSciInstrum}. 
The MIEZE method is particularly well suited to study dynamics close to the $[0,0,0]$ Bragg peak, corresponding to small-angle neutron scattering (SANS) \cite{2019_Franz_NuclInstrumMethodsPhysRes,2016_Krautloher_RevSciInstrum}. 
In \fig\ref{fig:fig_app_1_setup_masks} the experimental setup used for our experiments is shown.
The sample-detector distance, $L_{\mathrm{SD}}\,=\,\SI{2.335}{m}$, was maximized to ensure highest $q$ resolution. 
To provide high neutron flux and to cover the desired dynamic range, the wavelength was set to \SI{6}{\angstrom} with a wavelength spread $\Delta\lambda/\lambda\,=\,0.12$. 
In this configuration, a dynamic range from $\sim\,6\cdot10^{-6}\,\si{ns}$ to \SI{2}{ns} was accessible. 
\fig\ref{fig:fig_app_1_setup_masks}(b) shows a schematic of the neutron flight path through the spectrometer.
Data were recorded with a \SI{20}{cm}$\,\times\,$\SI{20}{cm} 2D CASCADE detector\cite{2011_Haussler_RevSciInstrum},  covering a $q$ range from \SI{0.016}{\angstrom^{-1}} to \SI{0.085}{\angstrom^{-1}} at $\lambda\,=\,$ \SI{6}{\angstrom} and $L_{\mathrm{SD}}\,=\,\SI{2.335}{m}$. Recent developments at RESEDA made it possible to increase $L_{\mathrm{SD}}$ up to \SI{3.43}{m} improving the spatial resolution further \cite{2019_Franz_NuclInstrumMethodsPhysRes}.

The grouping of the detector segments for evaluating the quasielastic data are shown in \fig\ref{fig:fig_app_1_setup_masks}(c). 
For the elastic measurements, the identical setup was used with narrower grouped detector segments, \ie data were evaluated in the same area as in \fig\ref{fig:fig_app_1_setup_masks}(c) but divided in 25 instead of 11 detector segments.
The sample was cooled to temperatures between \SI{4}{\kelvin} and \SI{300}{\kelvin} using a top-loading closed-cycle refrigerator and the temperature was controlled with two sensors close to the sample.
The temperature stability was $\sim$\SI{0.05}{\kelvin}. No hysteresis was observed in temperature scans.
Data were normalized using data recorded at the base temperature of $\sim$\SI{4}{\kelvin}, assuming that the spin dynamics in \ce{Fe_{x}Cr_{1-x}} are frozen at temperatures well below $T_{\mathrm{g}}$ \cite{2003_Pappas_PhysRevB}. 
This procedure minimizes systematic errors since the experimental setup remains unchanged throughout the entire experiment on a given sample.

We investigated three textured polycrystalline samples of \ce{Fe_{x}Cr_{1-x}} containing large grains with iron concentrations of $x\,=\,0.145$, $0.175$, and $0.21$. 
The samples were prepared by means of arc melting from pure starting materials and annealed for \SI{4}{days} at \SI{1100}{\celsius} before quenching in water \cite{1981_Shapiro_PhysRevB}.  
To remove strain, the samples were subsequently annealed at \SI{1000}{\celsius} for \SI{1}{day} \cite{1981_Shapiro_PhysRevB}. 
Following this process ingots of cylindrical shape with a height of approximately \SI{20}{mm} and a diameter of about \SI{10}{mm} were obtained.

To optimize the signal-to-noise ratio, slabs with a thickness of \SI{8}{mm} were cut from the ingots. A circular aperture made of cadmium with a diameter of \SI{10}{mm} was attached directly to the samples. Thus, the neutron beam effectively illuminated samples of cylindrical shape with a diameter of \SI{10}{mm} and a length of \SI{8}{mm}, where the cylindrical axis was parallel to the incident neutron beam.
Measurements on a sample with $x\,=\,0.17$ from Benka \etal \cite{2022_Benka_PhysRevMaterials} (not shown) were in excellent agreement with the results presented in the following.

Parts of the sample with $x\,=\,0.145$ were recently used to investigate the influence of concentration fluctuations on relaxation processes in spin-glasses \cite{2018_Wagner_QuantumBeamSci}. 
Using atom probe tomography, a high-resolution local probe, together with neutron resonant spin-echo spectroscopy it was shown that small-scale inhomogenieties in the microstructure influence the relaxation processes of a spin-glass material \cite{2018_Wagner_QuantumBeamSci}. 
Using the Weron model which had previously been applied to dilute spin-glasses \cite{2009_Pickup_PhysRevLett}, the relaxation processes were described in this study.

\begin{figure}[b]
    \centering
    \includegraphics[width=1.0\columnwidth]{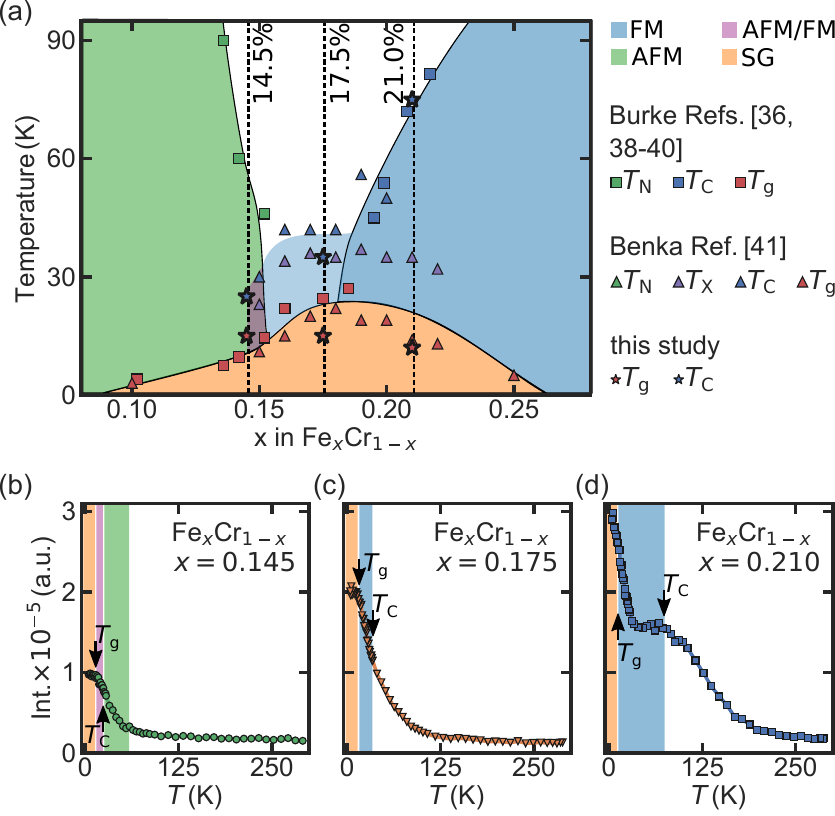}
    \caption{\captionheader{Magnetic properties of the \ce{Fe_{x}Cr_{1-x}} system. Antiferromagnetic (AFM, green), ferromagnetic (FM, blue), spin-glass (SG, orange), and antiferromagnetic-ferromagnetic (AFM/FM, purple) regimes are distinguished.} 
    (a) Temperature vs concentration phase diagram combining data from neutron scattering and low-field magnetization by Burke \etal (squares) \cite{1978_Burke_JPhysFMetPhys,1983_Burke_JPhysFMetPhysb,1983_Burke_JPhysFMetPhys,1983_Burke_JPhysFMetPhysa}, ac susceptibility and magnetization by Benka \etal (triangles) \cite{2022_Benka_PhysRevMaterials}, and neutron scattering (present work, stars). 
    (b)-(d) Integrated SANS intensity for the three samples measured at RESEDA. The shaded areas indicate magnetic regimes as inferred from the phase diagram.}
    \label{fig:fig_1_tscan_pd}
\end{figure}

\subsection{Experimental Results}
\label{sec:experimental-results}

The phase diagram of \ce{Fe_{x}Cr_{1-x}} as a function of temperature and iron concentration $x$ is shown in \fig\ref{fig:fig_1_tscan_pd}(a) as reproduced from literature \cite{1978_Burke_JPhysFMetPhys,1983_Burke_JPhysFMetPhysb,1983_Burke_JPhysFMetPhys,1983_Burke_JPhysFMetPhysa,2022_Benka_PhysRevMaterials}. 
For comparison, the temperature dependence of the integrated SANS intensity of the three samples measured in this study is shown in \figs\ref{fig:fig_1_tscan_pd}(b) to \ref{fig:fig_1_tscan_pd}(d). 
At the border between ferromagnetic and antiferromagnetic order, a dome of spin-glass behavior emerges, covering the regime of putative quantum phase transitions. 
At low temperatures, this spin-glass regime includes concentrations for which at high temperatures long-range ferromagnetic or antiferromagnetic order are observed.

In our study, three compositions were investigated: (i) $x\,=\,0.145$ which exhibits transitions from paramagnetism (PM) to antiferromagnetism (AFM) to a spin-glass (SG) with a glass temperature $T_g\,=\,\SI{11}{K}\pm\SI{2}{K}$ \cite{2022_Benka_PhysRevMaterials}, (ii) $x\,=\,0.175$ which exhibits transitions from PM via a small region reminiscent of FM order \cite{2022_Benka_PhysRevMaterials} to a SG with a glass temperature $T_g\,=\,\SI{20}{K}\pm\SI{2}{K}$ \cite{2022_Benka_PhysRevMaterials}, and (iii) $x\,=\,0.21$ which exhibits transitions from PM to ferromagnetism (FM) to a SG with a glass temperature $T_g\,=\,\SI{14}{K}\pm\SI{2}{K}$ \cite{2022_Benka_PhysRevMaterials}. 

\subsubsection{Elastic Scattering}
\label{sub:elastic-scattering-fecr}

The temperature dependence of the integrated SANS intensity is shown in \figs\ref{fig:fig_1_tscan_pd}(b) to \ref{fig:fig_1_tscan_pd}(d). 
As no magnetic scattering was observed at high temperatures, the data recorded at $\sim$\SI{300}{K} were used for background subtraction. 

The scattered intensities as a function of temperature for $x\,=\,0.145$ and $x\,=\,0.175$ behave similarly. 
With decreasing temperature the intensity increases as expected for a transition into a ferromagnetically ordered state.
When entering the spin-glass state, the system becomes static on the time scales probed by SANS, and the intensity increases with a change in slope, forming a plateau that starts at the onset of the spin-glass regime \cite{Mirebeau:uq5002}. 

The bulk properties and phase diagram show that the sample with $x\,=\,0.145$ enters the spin-glass regime via an antiferromagnetic state, which may not be identified microscopically for the parameter range probed in SANS. 
Interestingly, the increase in intensity is, analogous to the sample with $x\,=\,0.175$, reminiscent of ferromagnetic order in agreement with \reftaken\citen{2022_Benka_PhysRevMaterials}. 

For the sample with $x\,=\,0.21$, a broad feature with a maximum at $\sim$\SI{75}{\kelvin} defines $T_{\mathrm{C}}$, followed by a sharp increase in intensity. 
A change of slope with decreasing temperature, close to the transition temperature reported previously in \reftaken\citen{2022_Benka_PhysRevMaterials}, defines $T_{\mathrm{g}}$.
The signatures defined in neutron scattering are denoted by stars in \fig\ref{fig:fig_1_tscan_pd}(a). 
Discrepancies as compared to the phase boundaries inferred from the ac susceptibility and magnetization \cite{2022_Benka_PhysRevMaterials} may reflect the different time scales probed by the different methods.

\begin{figure}[ht]
    \centering
    \includegraphics[width=1.0\columnwidth]{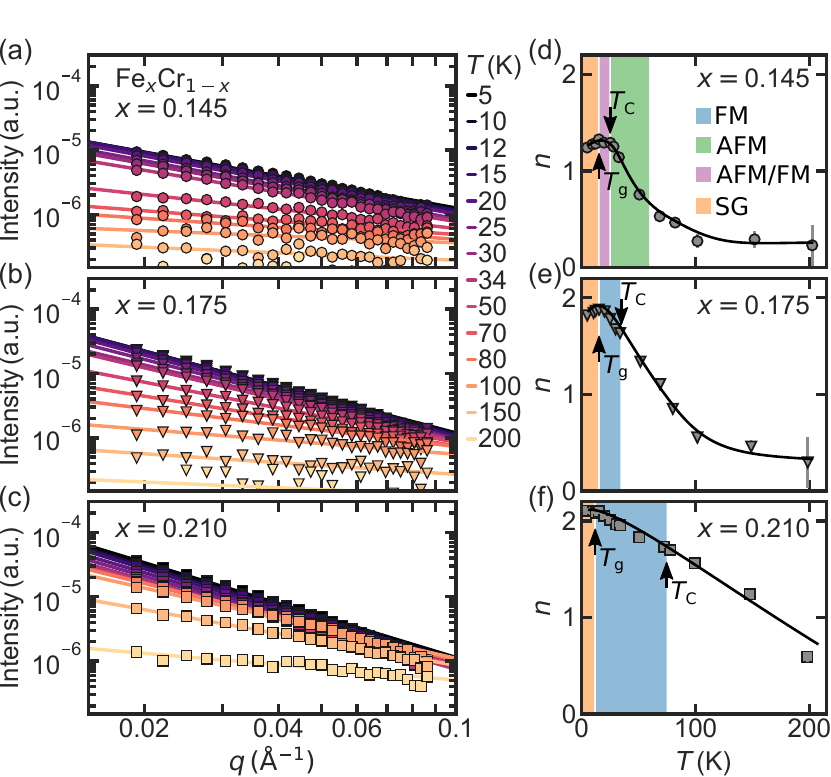}
    \caption{\captionheader{Momentum dependence of SANS of \ce{Fe_{x}Cr_{1-x}} as measured at RESEDA.}  
    (a)-(c) Intensity as a function of scattering vector $q$. The accessible $q$ range was $0.02\,\si{\angstrom^{-1}} < q < 0.08\,\si{\angstrom^{-1}}$. Solid lines are fits using \eq{\ref{eq:Iqn-sans}}.
    (d)-(f) Temperature dependence of the exponent $n$ obtained from the fits in (a)-(c). Solid lines are guides to the eye. The shaded areas indicate the regimes according to the phase diagram, namely AFM (green), AFM/FM (purple), FM (blue), and SG (orange). Samples with iron concentrations of $x=0.145$ (top row), $x=0.175$ (middle row), and $x=0.21$ (bottom row) were evaluated at different temperatures.
    }
    \label{fig:fig_app_2_IvsQ_nvsT}
\end{figure}

The $q$ dependence of the SANS data, \cf \sect\ref{sec:experiment} for details on data analysis, is shown in \figs\ref{fig:fig_app_2_IvsQ_nvsT}(a) to \ref{fig:fig_app_2_IvsQ_nvsT}(c) for different temperatures. 
Following the approach taken in related SANS studies on other materials \cite{2003_Mercone_PhysRevB,2004_Viret_PhysRevLett} we consider a single power law form
\begin{align}
    \label{eq:Iqn-sans}
    I \,\propto\, q^{-n}.
\end{align}
Fitting the experimental data yields exponents as a function of temperature as shown in \figs\ref{fig:fig_app_2_IvsQ_nvsT}(d) to \ref{fig:fig_app_2_IvsQ_nvsT}(f). 

With decreasing temperature, the exponent $n$ increases from $\sim\,0.3$, reaching low-temperature values of $n\,\sim\,1.3$ for $x\,=\,0.145$, $n\,\sim\,1.8$ for $x\,=\,0.175$, and $n\,\sim\,2.1$ for $x\,=\,0.21$. 
Even though the data are well-described by the power law in \eq{\ref{eq:Iqn-sans}}, unambiguous interpretation of the exponent $n$ proves difficult.
Assuming a two-phase system where smooth clusters are isolated, $n\,=\,4$ would be expected. Deviations from this behavior would indicate more complex phases including fractal surfaces. 
An exponent of $n\,=\,2$, for instance, was reported in the perovskite manganite \ce{Pr_{1-x}Ca_{x}MnO3} and attributed to sheets of inter-penetrating ferromagnetic and antiferromagnetic phases \cite{2003_Mercone_PhysRevB}. 
In the same material, Viret \etal  \cite{2004_Viret_PhysRevLett} found $n\,=\,1.6-1.7$ which was attributed to filamentary ferromagnetic chains in analogy to polymers.
An increase in $n$ with decreasing temperature suggests a coarsening of the interfaces as the system enters the spin-glass regime, consistent with an increase in cluster size. 

To address the question of the nature of the low-temperature magnetic structure as a function of iron concentration, it may be necessary to collect data at even smaller $q$ values, which is beyond the technical limits of the work reported here. Alternatively, more complex dependencies, \eg comprising several different power-law contributions may be possible. Such descriptions would require theoretical modelling beyond the scope of our study. 

\subsubsection{Quasielastic Measurements} 
\label{sub:quasielastic-measurements-fecr}

\begin{figure*}[tb]
	\centering
	\includegraphics[width=1.0\textwidth]{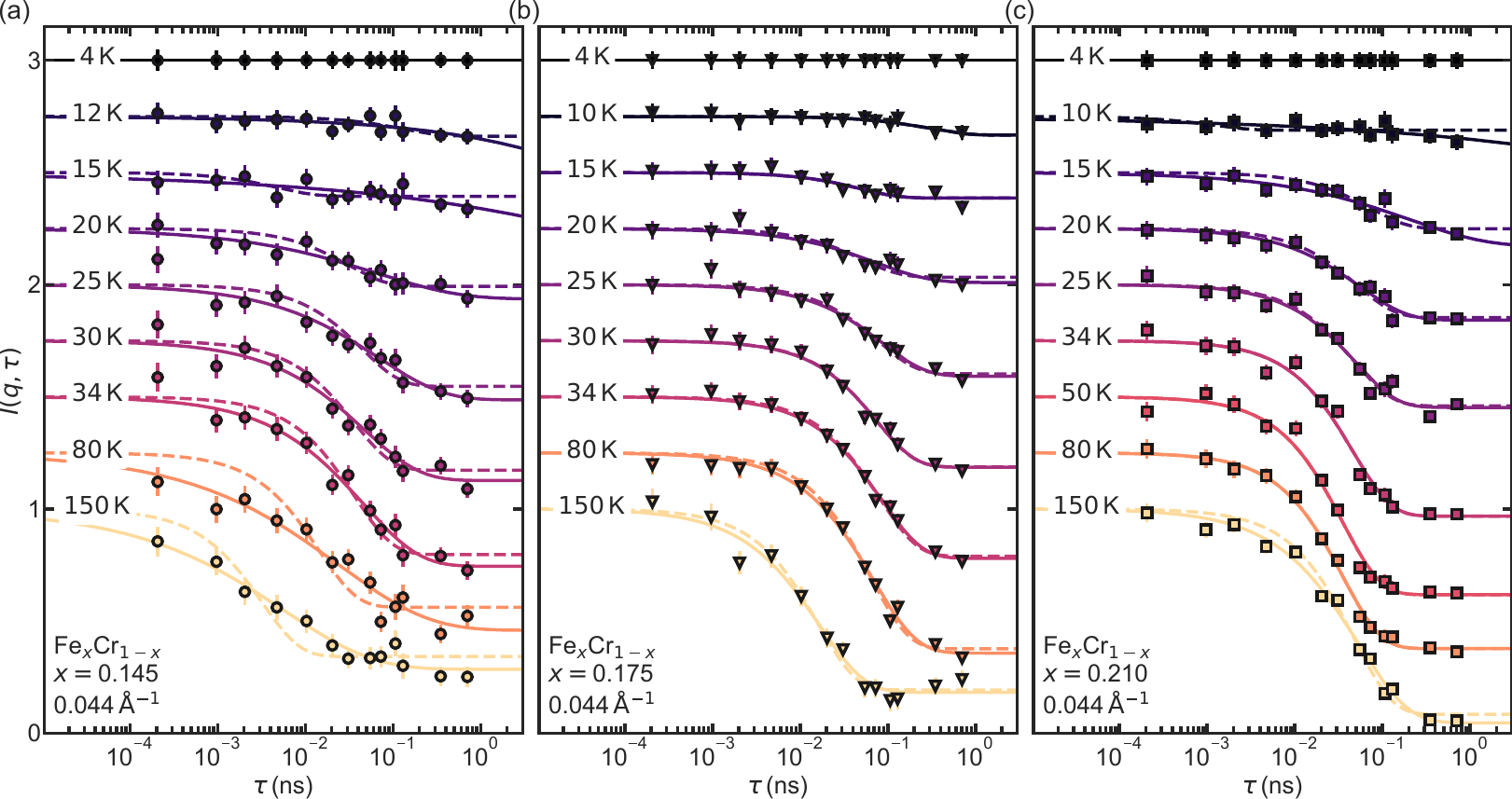}
	\caption{\captionheader{Normalized intermediate scattering function as measured for different temperatures in \ce{Fe_{x}Cr_{1-x}}.} Data were recorded using neutrons with a mean wavelength of $\lambda\,=\,6.0\,\si{\angstrom}$ and are shown at $q\,=\,0.044\,\si{\angstrom}^{-1}$ for (a) $x=0.145$, (b) $x=0.175$, (c) $x=0.21$. For better visibility, data are shifted vertically. The solid lines are fits to the data using the stretched exponential in \eq{\ref{eq:stretched-exp-decay}}. The dashed lines represent fits using a single exponential relaxation, corresponding to $\beta\,=\,1$ in \eq{\ref{eq:stretched-exp-decay}}. 
	}
	\label{fig:fig_3_mieze}
\end{figure*}

The normalized intermediate scattering function,
\begin{align}
    \label{eq:IqtSqt}
    I(q,\tau) = S(q,\tau)/S(q,0),
\end{align}
in the temperature range $\SI{4}{\kelvin} \,\leq\, T \,\leq\, \SI{150}{\kelvin}$ for $q\,=\,\SI{0.044}{\angstrom^{-1}}$ and all \ce{Fe_xCr_{1-x}} compositions investigated is shown in \fig\ref{fig:fig_3_mieze}. 
Following careful comparison of the data with different relaxation models models \cite{1979_Mezei_JMagnMagnMater,1981_Murani_JMagnMagnMater,1982_Mezei_JApplPhys,1983_Mezei_JMagnMagnMater,1985_Shapiro_JApplPhys,2003_Pappas_PhysRevB, 2009_Pickup_PhysRevLett,2007_Pickup_PhysicaB} we find that the data are already described well by a stretched exponential characteristic of a distribution of relaxation rates, namely 
\begin{align}
    \label{eq:stretched-exp-decay}
    I(q,\tau) \,=\, I_{\mathrm{elastic}} + (1-I_{\mathrm{elastic}}) \exp{\left(-\left(\Gamma \tau\right)^{\beta}\right)}.
\end{align}
Here, $\Gamma$ is the decay rate, corresponding to the inverse spin relaxation time, $0\,<\,\beta\,\leq\,1$ stretches the classic exponential decay ($\beta\,=\,1$), and the prefactor $I_{\mathrm{elastic}}$ describes the elastic contribution to the signal. In the light of the microscopic complexity of the Fe$_{x}$Cr$_{1-x}$ system, which may support both parallel as well as hierarchical relaxation, we refrain from an analysis in terms of more sophisticated mechanisms as assumed in the Weron model \cite{2018_Wagner_QuantumBeamSci,2009_Pickup_PhysRevLett,1991_Weron_JPhysCondensMatter,2007_Pickup_PhysicaB}.

The normalized intermediate scattering functions in \fig\ref{fig:fig_3_mieze} have been shifted vertically in steps of $0.25$ for better visibility. \fig\ref{fig:fig_3_mieze}(a) shows the sample with $x\,=\,0.145$. The intermediate scattering function is constant below $\sim$\SI{4}{\kelvin}, confirming that the spin dynamics are frozen \cite{2003_Pappas_PhysRevB} with respect to the normalizing data at \SI{4}{K}.
For temperatures around the glass temperature, $T_{\mathrm{g}}\,\sim\,\SI{15}{\kelvin}$, the sample starts to show signatures of dynamic behavior, notably spin relaxation as indicated by a decrease in the intermediate scattering function. 
The relaxation time of the spins is inferred from a fit to the data using $\tau = \frac{\hbar}{\Gamma}$. 
Increasing the temperature shifts the spin relaxation to shorter spin-echo times.
$I_{\mathrm{elastic}}$ describes the elastic contribution to the intermediate scattering function, corresponding to the fraction of the sample that is static on the time scales probed here, \ie  clusters fluctuating on much longer time scales.

The elastic signal decreases with increasing temperature, reaching a minimum of $I_{\mathrm{elastic}}\,\approx\,0.25$ for temperatures above $\SI{34}{\kelvin}$. 
For all temperatures, the system may be described by a stretched exponential decay, $\beta\,<\,1$. 
To highlight the need for the stretching parameter $\beta$, exponential decays ($\beta\,=\,1$) are shown for comparison as dashed lines in \figs \ref{fig:fig_3_mieze}(a) to \ref{fig:fig_3_mieze}(c). 
The deviation from simple exponential behavior gets more pronounced with decreasing temperature. 

The intermediate scattering functions for the sample with $x\,=\,0.175$ are shown in \fig\ref{fig:fig_3_mieze}(b). 
For temperatures below $T_{\mathrm{g}}$, the data resemble the behavior observed for $x\,=\,0.145$. Under increasing temperature the spin relaxation is similarly shifted to shorter times but remains larger than for $x\,=\,0.145$. 
The minimum of the elastic background is reduced to $I_{\mathrm{elastic}}\,\approx\,0.15$. 
In contrast to the sample with $x\,=\,0.145$, at higher temperatures the spin relaxation can be described by a simple exponential decay. 

For $x\,=\,0.21$, shown in \fig\ref{fig:fig_3_mieze}(c), the behavior again is highly reminiscent of the other two compositions. Accordingly, we observe a shift of the relaxation time towards shorter times as the temperature increases. 
However, for temperatures above \SI{50}{K} the relaxation time no longer decreases.
The elastic contribution reaches a minimum of $I_{\mathrm{elastic}}\,\approx\,0.1$. 
In a paramagnetic sample, where all scattering is dynamic, the spin-echo curve would decay to $I_{\mathrm{elastic}}\,=\,0$ at high temperatures or long spin-echo times. 
The observed sample dependence of $I_{\mathrm{elastic}}$ may be attributed empirically to the different iron contents of the samples, which lead to different high-temperature magnetic phases.

Since the measurements were performed in a SANS geometry, close to the ferromagnetic Bragg peak at $q\,=\,0$, ferromagnetic fluctuations will contribute strongly to the measured intensity. 
However, as possible antiferromagnetic scattering intensity cannot be observed in the vicinity of $q\,=\,0$, antiferromagnetic fluctuations will not contribute to the exponential decay, decreasing the dynamic contribution to the intensity in samples with a larger antiferromagnetic fraction. 
This observation is in accordance with the elastic measurements shown in \figs\ref{fig:fig_1_tscan_pd}(b) to \ref{fig:fig_1_tscan_pd}(d), which indicate that the intensity increases with increasing iron content. 

\begin{figure}[tb]
	\centering
	\includegraphics[width=1.0\columnwidth]{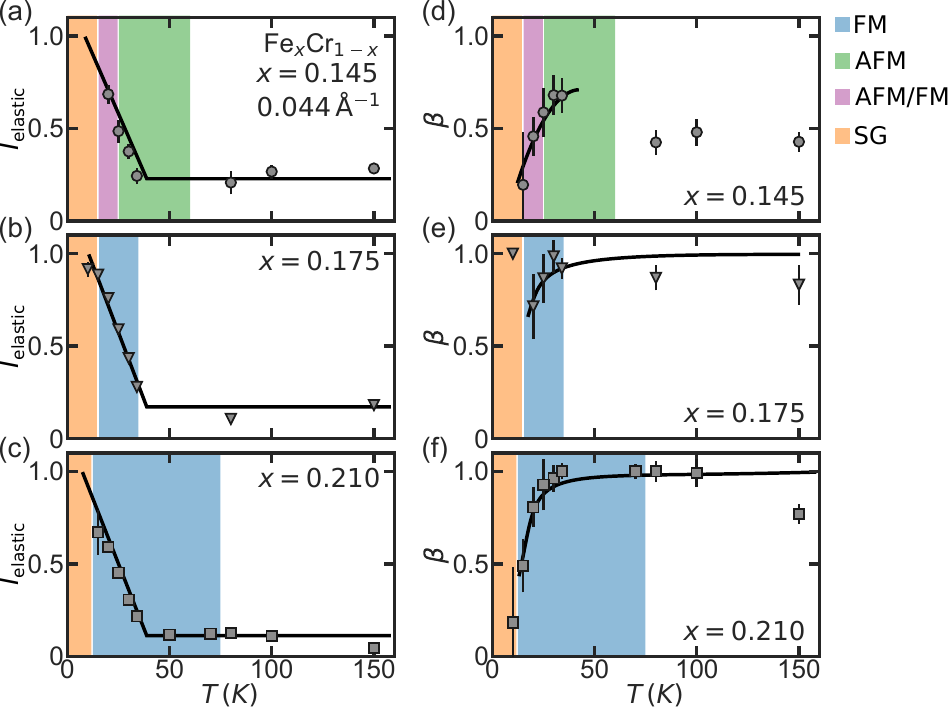}
	\caption{\captionheader{Temperature dependence of the fit parameters $I_{\mathrm{elastic}}$ and $\beta$ in \eq{\ref{eq:stretched-exp-decay}} for the three \ce{Fe_xCr_{1-x}} samples with (a),(d) $x\,=\,0.145$, (b),(e) $x\,=\,0.175$, and (c),(f) $x\,=\,0.21$. Data are shown for $q\,=\, \SI{0.044}{\angstrom^{-1}}$. The shaded areas indicate the phases according to the phase diagram: AFM (green), AFM/FM (purple), FM (blue), and SG (orange).} 
    (a)-(c) Elastic contribution $I_{\mathrm{elastic}}$ as a function of temperature, showing a linear increase for temperatures below $\sim$\SI{40}{K}. Solid lines are guides to the eye. The elastic background, \ie the constant value at temperatures above \SI{40}{K}, is connected to the magnetic signal-to-noise ratio, and thus different for the different samples, see main text for details.
    (d)-(f) Stretched exponent $\beta$ as function of temperature showing a drastic decrease of $\beta$ close to $T_{\mathrm{g}}$. Solid lines are guides to the eye.}
	\label{fig:fig_4_fit_Abeta}
\end{figure}

The temperature dependences of the fit parameters $I_{\mathrm{elastic}}$ and $\beta$ are summarized in \fig\ref{fig:fig_4_fit_Abeta} for $q$\,=\,\SI{0.044}{\angstrom^{-1}}. 
For all samples the elastic contribution $I_{\mathrm{elastic}}$ decreases linearly with increasing temperature reaching a constant value above $\sim$\SI{40}{K}, \ie far above the glass temperature. 
A shrinking elastic contribution with increasing temperature suggests that parts of the sample slowly unfreeze on the time scales studied in our experiments. 

The exponent $\beta$ provides an estimate of the broadening of the spectrum of relaxation times. 
For all samples, $\beta$ decreases drastically when the temperature decreases towards the spin-glass regime. This evolution suggests an increase of the distribution of relaxation times, as expected of the formation of different-sized domains fluctuating on different time scales before freezing at the lowest temperatures. 
For $x\,=\,0.145$, $\beta$ stays below 1 over the entire temperature range, while for the other two compositions $\beta$ is close to 1 at temperatures above $T_{\mathrm{g}}$. This means that the dynamics may be described with a single relaxation time.

Numerical calculations of the non-exponential relaxation in spin-glasses and glassy systems \cite{1987_Campbell_JPhysCSolidStatePhys,1988_Campbell_PhysRevB} have shown that $\beta\,=\,1/3$ is approached at $T_{\mathrm{g}}$ when the system is close to its percolation limit. 
The decrease of $\beta$ we observe as the spin-glass regime is approached is in qualitative agreement with these calculations.

\subsection{\label{sec:discussion}Discussion}

\begin{figure}[tb]
    \centering
    \includegraphics[width=1.0\columnwidth]{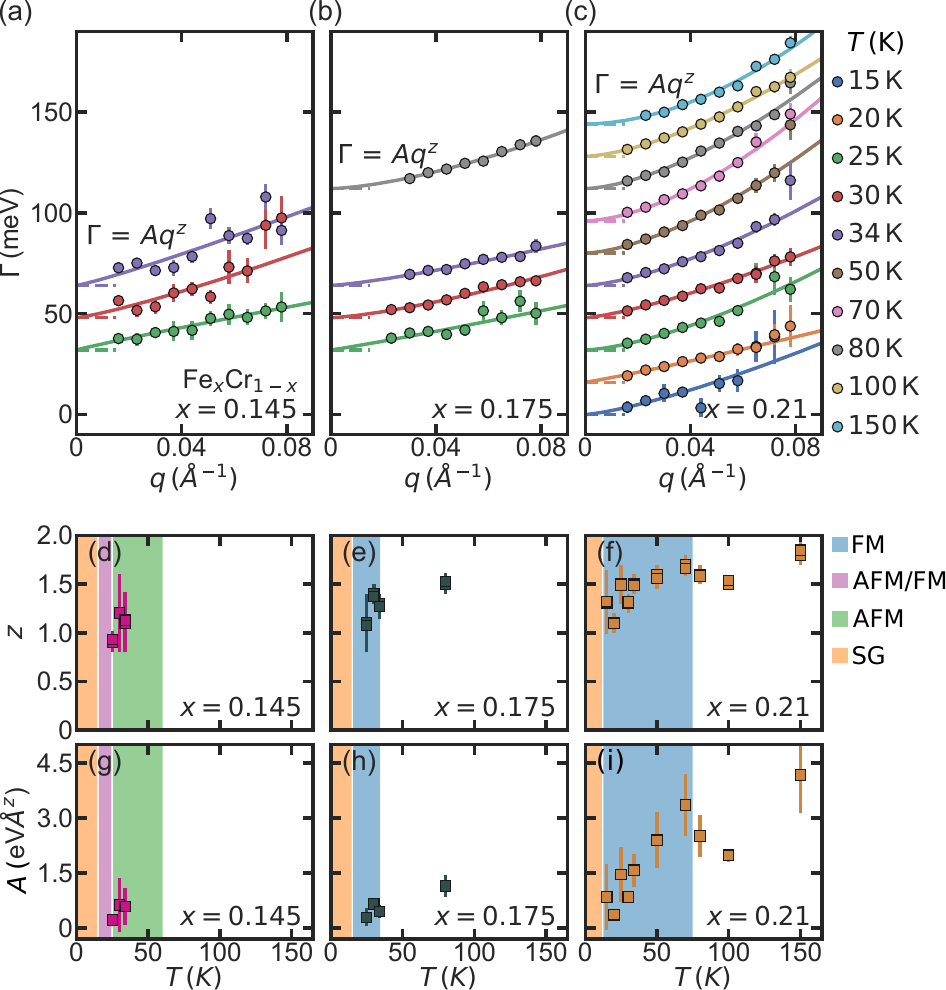}
    \caption{\captionheader{Momentum dependence of the decay rate $\Gamma$.} 
    Data are shown for (a) $x\,=\,0.145$, (b) $x\,=\,0.21$, and (c) $x\,=\,0.175$ and temperatures where a meaningful linewidth $\Gamma$ could be extracted, see main text for details. The solid lines are fits to the data using \eq{\ref{eq:gammaqz-fecr}}. The temperature dependence of the fit parameters $z$ and $A$ are shown for $x\,=\,0.145$ (left column), $x\,=\,0.175$ (middle column), and $x\,=\,0.21$ (right column). The shaded areas indicate the phases according to the phase diagram: AFM (green), AFM/FM (purple), FM (blue), and SG (orange).}
    \label{fig:fig_5_GvsQ}
\end{figure}

The area detector used in our MIEZE measurements  
allowed studying the normalized intermediate scattering function $I(q,\tau)$ simultaneously over a wide range of momentum transfers $q$. The $q$ dependence of the decay rate $\Gamma$ and therefore the spin relaxation time is shown in \figs\ref{fig:fig_5_GvsQ}(a) to \ref{fig:fig_5_GvsQ}(c). Within experimental accuracy, our data it are consistent with 
\begin{align}
    \label{eq:gammaqz-fecr}
    \Gamma = A q^z,
\end{align}
where $q$ is the momentum transfer, $z$ the dynamical exponent, and $A$ the energy scale of the exchange interactions.

The decay rate $\Gamma$ as a function of $q$, shown in \fig\ref{fig:fig_5_GvsQ}, was fitted using \eq{\ref{eq:gammaqz-fecr}} with  fixed values of $z$, \ie $z\,=\,1.0$, $z\,=\,1.5$, and $z\,=\,2.0$, as well as with $z$ as an independent fitting parameter. 
Additionally, data were fitted for each sample and for all temperatures independently, as well as simultaneously for all temperatures. 
The fit results are summarized in \tab\ref{tab:GvsQfit}. 
A $\chi^2$-analysis of the fits with fixed parameter $z$ shows that with decreasing iron content $z$ decreases from $\sim1.5$ to $\sim1.0$, which is supported by the fits with $z$ as a free parameter. 
The resulting values of $z$ and $A$ as a function of temperature are depicted in \figs\ref{fig:fig_5_GvsQ}(d) to \ref{fig:fig_5_GvsQ}(i).

For small $\beta$, the distribution of relaxation times is very broad and the data cannot be described with a single $\tau$. 
Therefore a meaningful linewidth $\Gamma$ cannot be extracted for any of the three samples below and around the glass temperature $T_{\mathrm{g}}$.
For $x\,=\,0.145$, $\Gamma$ could only be extracted for a small temperature window in the ferromagnetic regime, $\SI{25}{\kelvin}\,\leq\,T\,\leq\,\SI{34}{\kelvin}$. 
Even at these temperatures, the exponential decay is already stretched and the values of $\Gamma$ determined experimentally represent a mean relaxation time rather than a single relaxation time. 
At higher temperatures, the exponential decay is strongly stretched such that a single dominant relaxation time could not be determined. 
The sample with $x\,=\,0.175$ allowed us to analyze $\Gamma$ for temperatures between $\SI{25}{\kelvin}\,\leq\,T\,\leq\,\SI{80}{\kelvin}$. For $x\,=\,0.21$, $\Gamma$ could be extracted for all temperatures above the freezing temperature $T_{\mathrm{g}}$.

\begin{table*}[htbp]
\caption{Summary of the fitting procedure of the spin dynamics in \ce{Fe_xCr_{1-x}} using \eq{\ref{eq:gammaqz-fecr}} with different approaches, namely: (i) fixing the exponent $z\,=\,1.0$, (ii) fixing the exponent $z\,=\,1.5$, (iii) fixing the exponent $z\,=\,2.0$, (iv) leaving the exponent $z$ as a free fit parameter. Additionally, data were also fitted with (I) all temperatures independently and (II) all temperatures simultaneously. The $\chi^2$ value of each fit is used as an indicator for the goodness of the fits.}
\label{tab:GvsQfit}
\begin{tabular}{|l||l|l|l|l|l||l|l|l|l|l||l|l|l|l|l||l|l|l|l|l|}
\multicolumn{21}{c}{\bf\ce{Fe_xCr_{1-x}} with $x=0.145$; all temperatures fitted independently}\\
\hline
& \multicolumn{5}{c}{$z\,=\,1.0$}   & \multicolumn{5}{|c}{$z\,=\,1.5$}        & \multicolumn{5}{|c}{$z\,=\,2.0$}        & \multicolumn{5}{|c|}{$z\,=\,$free}        \\ \hline
$T\,$(K)  & $A$    & $A_{\mathrm{err}}$ & $z$   & $z_{\mathrm{err}}$ & $\chi^2$   & $A$    & $A_{\mathrm{err}}$ & $z$   & $z_{\mathrm{err}}$ & $\chi^2$   & $A$    & $A_{\mathrm{err}}$ & $z$   & $z_{\mathrm{err}}$ & $\chi^2$   & $A$    & $A_{\mathrm{err}}$ & $z$   & $z_{\mathrm{err}}$ & $\chi^2$   \\ \hline
25 & 273 &  10 & 1.0 &   $-$ &  1.20  & 1127 &  84 & 1.5 &   $-$ &  4.95  & 4394 &  506 & 2.0 &   $-$ & 11.19 & 223 &  61 & 0.9 &   0.1 &  1.13 \\
30 & 333 &  48 & 1.0 &   $-$ & 18.63  & 1536 & 225 & 1.5 &   $-$ & 18.84  & 6576 & 1042 & 2.0 &   $-$ & 21.47 & 629 & 745 & 1.2 &   0.4 & 18.22 \\
34 & 400 &  40 & 1.0 &   $-$ & 20.21  & 1723 & 184 & 1.5 &   $-$ & 22.79  & 6953 &  914 & 2.0 &   $-$ & 32.99 & 589 & 507 & 1.1 &   0.3 & 19.74 \\
\hline
\multicolumn{21}{c}{}\\
\multicolumn{21}{c}{\bf\ce{Fe_xCr_{1-x}} with $x=0.145$; all temperatures fitted simultaneously}\\
\hline
   & \multicolumn{5}{c}{$z\,=\,1.0$}   & \multicolumn{5}{|c}{$z\,=\,1.5$}        & \multicolumn{5}{|c}{$z\,=\,2.0$}        & \multicolumn{5}{|c|}{$z\,=\,$free}        \\ \hline
$T\,$(K)  & $A$    & $A_{\mathrm{err}}$ & $z$   & $z_{\mathrm{err}}$ & $\chi^2$   & $A$    & $A_{\mathrm{err}}$ & $z$   & $z_{\mathrm{err}}$ & $\chi^2$   & $A$    & $A_{\mathrm{err}}$ & $z$   & $z_{\mathrm{err}}$ & $\chi^2$   & $A$    & $A_{\mathrm{err}}$ & $z$   & $z_{\mathrm{err}}$ & $\chi^2$   \\ \hline
all &  336 &  22 & 1.0 &   $-$ & 51.27  & 1443 & 104 & 1.5 &   $-$ & 60.63 & 5794 & 498 & 2.0 &   $-$ &  82.4 & 377.0 & 190.0 & 1.0 &   0.2 & 51.18 \\
\hline
\multicolumn{21}{c}{}\\
\multicolumn{21}{c}{}\\
\multicolumn{21}{c}{}\\
\multicolumn{21}{c}{\bf\ce{Fe_xCr_{1-x}} with $x=0.175$; all temperatures fitted independently}\\
\hline
   & \multicolumn{5}{c}{$z\,=\,1.0$}   & \multicolumn{5}{|c}{$z\,=\,1.5$}        & \multicolumn{5}{|c}{$z\,=\,2.0$}        & \multicolumn{5}{|c|}{$z\,=\,$free}        \\ \hline
$T\,$(K)  & $A$    & $A_{\mathrm{err}}$ & $z$   & $z_{\mathrm{err}}$ & $\chi^2$   & $A$    & $A_{\mathrm{err}}$ & $z$   & $z_{\mathrm{err}}$ & $\chi^2$   & $A$    & $A_{\mathrm{err}}$ & $z$   & $z_{\mathrm{err}}$ & $\chi^2$   & $A$    & $A_{\mathrm{err}}$ & $z$   & $z_{\mathrm{err}}$ & $\chi^2$   \\ \hline
25 & 233 &  17 & 1.0 &   $-$ &  7.60 & 1073 &  88 & 1.5 &   $-$ &  9.88 & 4663 & 537 & 2.0 &   $-$ & 18.63 &  293 & 249 & 1.1 &   0.3 &  7.54 \\
30 & 212 &  12 & 1.0 &   $-$ & 12.51 &  950 &  29 & 1.5 &   $-$ &  3.94 & 3912 & 306 & 2.0 &   $-$ & 24.21 &  656 & 158 & 1.4 &   0.1 &  2.94 \\
34 & 196 &   7 & 1.0 &   $-$ &  5.55 &  891 &  31 & 1.5 &   $-$ &  5.03 & 3877 & 273 & 2.0 &   $-$ & 20.56 &  445 & 178 & 1.3 &   0.1 &  3.34 \\
80 & 239 &  13 & 1.0 &   $-$ & 15.82 & 1060 &  22 & 1.5 &   $-$ &  2.30 & 4484 & 213 & 2.0 &   $-$ & 12.26 & 1145 & 298 & 1.5 &   0.1 &  2.26 \\
\hline
\multicolumn{21}{c}{}\\
\multicolumn{21}{c}{\bf\ce{Fe_xCr_{1-x}} with $x=0.175$; all temperatures fitted simultaneously}\\
\hline
   & \multicolumn{5}{c}{$z\,=\,1.0$}   & \multicolumn{5}{|c}{$z\,=\,1.5$}        & \multicolumn{5}{|c}{$z\,=\,2.0$}        & \multicolumn{5}{|c|}{$z\,=\,$free}        \\ \hline
$T\,$(K)  & $A$    & $A_{\mathrm{err}}$ & $z$   & $z_{\mathrm{err}}$ & $\chi^2$   & $A$    & $A_{\mathrm{err}}$ & $z$   & $z_{\mathrm{err}}$ & $\chi^2$   & $A$    & $A_{\mathrm{err}}$ & $z$   & $z_{\mathrm{err}}$ & $\chi^2$   & $A$    & $A_{\mathrm{err}}$ & $z$   & $z_{\mathrm{err}}$ & $\chi^2$   \\ \hline
all & 216 &   6 & 1.0 &   $-$ & 55.92  & 973 &  22 & 1.5 &   $-$ & 33.46 & 4138 & 149 & 2.0 &   $-$ & 86.81 & 700 & 163 & 1.4 &   0.1 & 31.52 \\
\hline
\multicolumn{21}{c}{}\\
\multicolumn{21}{c}{}\\
\multicolumn{21}{c}{}\\
\multicolumn{21}{c}{\bf\ce{Fe_xCr_{1-x}} with $x=0.21$; all temperatures fitted independently}\\
\hline
& \multicolumn{5}{c}{$z\,=\,1.0$}      & \multicolumn{5}{|c}{$z\,=\,1.5$}        & \multicolumn{5}{|c}{$z\,=\,2.0$}        & \multicolumn{5}{|c|}{$z\,=\,$free}        \\ \hline
$T\,$(K)  & $A$    & $A_{\mathrm{err}}$ & $z$   & $z_{\mathrm{err}}$ & $\chi^2$   & $A$    & $A_{\mathrm{err}}$ & $z$   & $z_{\mathrm{err}}$ & $\chi^2$   & $A$    & $A_{\mathrm{err}}$ & $z$   & $z_{\mathrm{err}}$ & $\chi^2$   & $A$    & $A_{\mathrm{err}}$ & $z$   & $z_{\mathrm{err}}$ & $\chi^2$   \\ \hline
 15 & 303 &  37 & 1.0 &   $-$ &   9.90  & 1495 & 173 & 1.5 &   $-$ &  9.14 & 6761 & 928 & 2.0 &   $-$ & 12.33 &  849 &  889 & 1.3 &   0.3 &  8.84 \\
 20 & 266 &   7 & 1.0 &   $-$ &   2.05  & 1204 &  61 & 1.5 &   $-$ &  6.80 & 5096 & 479 & 2.0 &   $-$ & 22.29 &  360 &   83 & 1.1 &   0.1 &  1.68 \\
 25 & 310 &  26 & 1.0 &   $-$ &  23.64  & 1504 &  83 & 1.5 &   $-$ & 10.65 & 6646 & 519 & 2.0 &   $-$ & 20.54 & 1462 &  742 & 1.5 &   0.2 & 10.64 \\
 30 & 311 &  15 & 1.0 &   $-$ &  11.96  & 1469 &  45 & 1.5 &   $-$ &  4.97 & 6321 & 511 & 2.0 &   $-$ & 33.39 &  844 &  149 & 1.3 &   0.1 &  2.26 \\
 34 & 338 &  24 & 1.0 &   $-$ &  29.58  & 1641 &  52 & 1.5 &   $-$ &  6.28 & 7232 & 457 & 2.0 &   $-$ & 24.47 & 1578 &  448 & 1.5 &   0.1 &  6.27 \\
 50 & 430 &  33 & 1.0 &   $-$ &  43.43  & 2011 &  66 & 1.5 &   $-$ &  8.49 & 8643 & 446 & 2.0 &   $-$ & 20.47 & 2393 &  764 & 1.6 &   0.1 &  8.20 \\
 70 & 418 &  37 & 1.0 &   $-$ &  43.72  & 2041 &  64 & 1.5 &   $-$ &  5.76 & 9099 & 392 & 2.0 &   $-$ & 10.77 & 3350 &  838 & 1.7 &   0.1 &  3.76 \\
 80 & 423 &  32 & 1.0 &   $-$ &  66.27  & 1952 &  48 & 1.5 &   $-$ &  7.52 & 8283 & 387 & 2.0 &   $-$ & 26.58 & 2506 &  565 & 1.6 &   0.1 &  6.48 \\
100 & 383 &  25 & 1.0 &   $-$ &  54.38  & 1763 &  18 & 1.5 &   $-$ &  1.30 & 7499 & 327 & 2.0 &   $-$ & 24.07 & 1984 &  181 & 1.5 &   0.1 &  1.07 \\
150 & 291 &  37 & 1.0 &   $-$ & 137.06  & 1457 &  78 & 1.5 &   $-$ & 27.54 & 6468 & 214 & 2.0 &   $-$ & 10.63 & 4169 & 1032 & 1.8 &   0.1 &  7.52 \\
\hline
\multicolumn{21}{c}{}\\
\multicolumn{21}{c}{\bf\ce{Fe_xCr_{1-x}} with $x=0.21$; all temperatures fitted simultaneously}\\
\hline
   & \multicolumn{5}{c}{$z\,=\,1.0$}   & \multicolumn{5}{|c}{$z\,=\,1.5$}        & \multicolumn{5}{|c}{$z\,=\,2.0$}        & \multicolumn{5}{|c|}{$z\,=\,$free}        \\ \hline
$T\,$(K)  & $A$    & $A_{\mathrm{err}}$ & $z$   & $z_{\mathrm{err}}$ & $\chi^2$   & $A$    & $A_{\mathrm{err}}$ & $z$   & $z_{\mathrm{err}}$ & $\chi^2$   & $A$    & $A_{\mathrm{err}}$ & $z$   & $z_{\mathrm{err}}$ & $\chi^2$   & $A$    & $A_{\mathrm{err}}$ & $z$   & $z_{\mathrm{err}}$ & $\chi^2$   \\ \hline
all &  348 &  10 & 1.0 &   $-$ & 610.23  & 1659 &  31 & 1.5 &   $-$ & 256.94 & 7203 & 163 & 2.0 &   $-$ & 363.71 & 2205 & 364 & 1.6 &   0.1 & 249.2 \\
\hline
\end{tabular}
\end{table*}

For all concentrations, $\Gamma$ depends on $q$ according to the relation given in \eq{\ref{eq:gammaqz-fecr}}. 
The exponent $z$ increases for increasing iron concentration from $z\,\approx\,1$ to $z\,\approx\,2$.
According to dynamic scaling theory the critical exponent at $T_{c}$ for pure ferromagnets in the limit $q\,\rightarrow\,0$ corresponds to $z\,=\,2.5$ while antiferromagnetic correlations for $q\,\rightarrow\,Q_{\mathrm{AFM}}$ lead to $z\,=\,1.5$.
Heuristically, one might hence explain the exponents in \ce{Fe_{x}Cr_{1-x}} in terms of a competition of ferromagnetic and antiferromagnetic correlations.

Tajima \etal used a spin diffusion model with $\Gamma \,\propto\, q^{2.0}$ to describe the $q$ dependence of $\Gamma$ in the Invar alloy \ce{Fe_{65}Ni_{35}} over a rather wide $q$ range \cite{1987_Tajima_PhysRevB}. 
This model assumes a hydrodynamic behavior of uncorrelated spins. The authors argued that the system never reaches criticality, \ie $z\,=\,2.5$ for ferromagnetically correlated spins, due to impurity scattering of the electrons. 

Along this line, the presence of chromium atoms acting as impurities in ferromagnetic iron clusters could prevent the system from reaching ferromagnetic critical dynamics, reducing $z$ to $2.0$.
A large fraction of chromium, and therefore an increase in antiferromagnetic correlations in \ce{Fe_{x}Cr_{1-x}} could reduce the exponent $z$ further towards $1.5$ leading to an effective range of exponents between $z\,=\,1.5$ for antiferromagnetic correlations and $z\,=\,2.0$ as expected in the spin diffusion model. 
Values of $z$ below $1$ may reflect the large number of different time scales in spin-glasses.
Unusually small values of $z$ found in spin-glasses were previously attributed to disorder, the proximity to the spin-glass transition or a reentrant phase, or the complex character of the interactions \cite{1992_Lartigue_PhysicaBCondensedMatter,1990_Pappas_PhysicaBCondensedMatter}.

A $q$ dependence of $\Gamma$ has, finally, been reported by Bao \etalk, who found that the spin dynamics in \ce{Fe_{0.7}Al_{0.3}} can be approximated with $\Gamma\,\propto\,q^{2.5}$ for $q$ values from $0.05\,$\AA$^{-1}\,-\,0.5$\,\AA$^{-1}$. 
This finding is in contrast to studies of dilute spin-glasses which reported that the relaxation process does not depend on $q$. Such a result is expected for dilute systems that are homogeneous over the $q$ scale studied. 

\subsection{\label{sec:conclusion}Conclusions}

In summary, we reported a MIEZE spectroscopy study of the spin dynamics in \ce{Fe_{x}Cr_{1-x}} samples with $x\,=\,0.145$, $x\,=\,0.175$, and $x\,=\,0.21$. 
These compositions were chosen to compare the spin relaxation for different spin-glass classifications and reentrant temperature dependencies from different high-temperature properties, namely antiferromagnetic, paramagnetic, and ferromagnetic states. The dynamic properties of all samples studied are consistent with a comprehensive study of polycrystaline samples \cite{2022_Benka_PhysRevMaterials}. Properties of the sample with $x\,=\,0.175$ have been reported in \reftaken\citen{2018_Wagner_QuantumBeamSci} using transverse NRSE and conventional NSE featuring more scatter and a smaller dynamic range. All samples show a broad distribution of relaxation times close to the spin-glass regime which, in the simplest approach, can be described by a stretched exponential.
The stretching exponent $\beta$ approaches $1/3$ as the spin-glass regime is approached, suggesting proximity  to the percolation limit. 
For increasing iron content and hence increasing tendency to ferromagnetic order, the spin relaxation can be described with a simple exponential (Debye) decay at high temperatures, where a single relaxation time can be extracted, indicating a smaller spread in relaxation times and magnetic cluster sizes. As explained above we refrain from an analysis in terms of more specific mechanisms as assumed, \eg in the Weron model.

The spin relaxation dynamics of all three compositions $x$ investigated in our study depends on $q$. Such a dependence contrasts dilute spin-glasses in which no dispersive behavior has been observed\cite{1984_Heffner_PhysRevB}. Within the scatter of our data the dispersive behavior may be described by a power-law dependence where the exponent $z$ decreases with increasing iron concentration. We note, however, that the dispersive character may be the result of a combination of different contributions with different values of $z$, a scenario that we cannot disentangle further.

On a technical note, the present study highlights that the MIEZE technique allows to perform high-resolution neutron spin-echo spectroscopy over a large dynamic range in materials hosting ferromagnetic domains that may depolarize the neutron beam, posing a major limitation in conventional neutron spin-echo spectroscopy. Samples with ferromagnetic, antiferromagnetic, or paramagnetic high-temperature properties may therefore be investigated using MIEZE with the same instrument and sample environment. 

\begin{acknowledgments}
We wish to thank F.~Haslbeck, M.~Mantwill, S.~Mayr, S.~Mühlbauer, and A.~Wendl for fruitful discussions and assistance with the experiments. This work has been funded by the Deutsche Forschungsgemeinschaft (DFG, German Research Foundation) under TRR80 (From Electronic Correlations to Functionality, Project No.\ 107745057, Project E1) and the excellence cluster MCQST under Germany's Excellence Strategy EXC-2111 (Project No.\ 390814868). Financial support by the Bundesministerium f\"{u}r Bildung und Forschung (BMBF) through Project No.\ 05K16WO6 as well as by the European Research Council (ERC) through Advanced Grants No.\ 291079 (TOPFIT) and No.\ 788031 (ExQuiSid) is gratefully acknowledged. G.B.\ and S.S.\ acknowledge financial support through the TUM Graduate School.
\end{acknowledgments}

\bibliographystyle{apsrev4-1edit}
\bibliography{library,library_stfn}
\end{document}